\documentclass[12pt,a4paper]{article}
\usepackage{amssymb,amsmath}
\usepackage{color} 
\usepackage{graphicx}

\newcommand{\bc}{\begin{center}}
\newcommand{\ec}{\end{center}}
\newcommand{\bd}{\begin{displaymath}}
\newcommand{\ed}{\end{displaymath}}
\newcommand{\be}{\begin{equation}}
\newcommand{\ee}{\end{equation}}
\newcommand{\ba}{\begin{array}}
\newcommand{\ea}{\end{array}}
\newcommand{\bt}{\begin{tabular}}
\newcommand{\et}{\end{tabular}}

\begin{document}

\title{Predicting the SUSY breaking scale in SUGRA models with degenerate vacua}

\author{C.~D.~Froggatt${}^{a}$,
R.~Nevzorov${}^{b}$,
H.~B.~Nielsen${}^{c}$,
A.~W.~Thomas${}^{d}$\\[5mm]
\itshape{$^a$ School of Physics and Astronomy, University of Glasgow,}\\[0mm]
\itshape{Glasgow, G12 8QQ, UK}\\[0mm]
\itshape{$^b$ NRC Kurchatov Institute --- ITEP, Moscow, 117218, Russia}\\[0mm]
\itshape{$^c$ The Niels Bohr Institute, University of Copenhagen,}\\[0mm]
\itshape{Blegdamsvej 17, Copenhagen, DK-2100, Denmark}\\[0mm]
\itshape{$^d$ ARC Centre of Excellence for Particle Physics at the Terascale and CSSM,}\\[0mm]
\itshape{Department of Physics, The University of Adelaide, Adelaide SA 5005, Australia}}

\date{}

\maketitle

\begin{abstract}{
\noindent
In $N=1$ supergravity the scalar potential may have supersymmetric (SUSY)
and non-supersymmetric Minkowski vacua (associated with supersymmetric and
physical phases) with vanishing energy density.
In the supersymmetric Minkowski (second) phase some breakdown of SUSY
may be induced by non-perturbative effects in the observable sector
that give rise to a tiny positive vacuum energy density.
Postulating the exact degeneracy of the physical and second vacua as well
as assuming that at high energies the couplings in both phases are almost
identical, one can estimate the dark energy density in these vacua.
It is mostly determined by the SUSY breaking scale $M_S$ in the physical phase. Exploring
the two-loop renormalization group (RG) flow of couplings in these vacua we find that the measured
value of the cosmological constant can be reproduced if $M_S$ varies from $20\,\mbox{TeV}$ to
$400\,\mbox{TeV}$. We also argue that this prediction for the SUSY breaking scale
is consistent with the upper bound on $M_S$ in the higgsino dark matter scenario.}
\end{abstract}

\newpage
\section{Introduction}

The tiny dark energy density (the cosmological constant $\rho_{\Lambda}$) spread
over the whole Universe, which is responsible for
its accelerated expansion, is a major puzzle for modern particle physics nowadays.
Its value, $\rho_{\Lambda} \sim 10^{-55} M_Z^4$,
is much smaller than the contribution to the cosmological constant that comes from the
electroweak (EW) symmetry breaking in the
standard model (SM), which should be of the order of $M_Z^4$, where $M_Z$ is the $Z$ boson mass.
At the same time the contribution of
zero--modes is expected to push the dark energy density even higher up to $\sim M_{P}^4$,
where $M_{P}$ is the Planck scale.
Because of the enormous cancellation between different contributions to $\rho_{\Lambda}$,
which would be required to keep the dark energy
density around its measured value, the smallness of the cosmological constant should be regarded as a fine--tuning problem.

At this moment none of the available extensions of the SM provides a satisfactory explanation for the smallness of the
dark energy density. In particular, in the models based on the local $(N=1)$ supersymmetry (SUSY), i.e. $(N=1)$ supergravity (SUGRA),
a huge degree of fine--tuning is needed to ensure that the energy density in the physical vacuum is sufficiently small.
Here we use the so-called Multiple Point Principle (MPP) \cite{mpp}-\cite{mpp-nonloc} to address the cosmological constant puzzle.
The MPP postulates the coexistence in Nature of many phases allowed by a given theory.
On the phase diagram of the theory
it corresponds to a special (multiple) point where these phases meet. At the multiple point the vacuum energy densities of
these different phases are degenerate. When applied to the SM, the MPP implies that the Higgs effective potential in this model
possesses two degenerate minima which are taken to be at the EW and Planck scales \cite{Froggatt(1996)}. This led to
the remarkable predictions for the pole masses of the top quark and Higgs boson,
i.e. $M_t=173\pm 5\,\mbox{GeV}$ and $M_H=135\pm 9\, \mbox{GeV}$.
The application of the MPP to the two Higgs doublet extension of the SM has also
been explored \cite{2hdm-1}--\cite{2hdm-2}. It was argued
that in this case the MPP can be used as a mechanism for the suppression of
CP--violation and flavour changing neutral currents \cite{2hdm-2}.

The application of the MPP to $(N=1)$ SUGRA models was considered as well \cite{Froggatt:2003jm}--\cite{Froggatt:2014jza}.
The successful implementation of the MPP in these models requires the presence of a supersymmetric Minkowski (second) vacuum.
This second vacuum and the physical one, in which we live, must have the same energy densities. The breakdown of local SUSY
in $(N=1)$ SUGRA models takes place in the hidden sector. This sector involves superfields $(z_i)$ that interact with
the observable ones only by means of gravity. The existence of the second vacuum implies that the superpotential $W(z_i)$
of the hidden sector has a stationary point, where it vanishes. The K$\Ddot{a}$hler potential and superpotential
of the simplest SUGRA model that satisfies this condition can be written as \cite{Froggatt:2003jm}
(for recent review see \cite{Froggatt:2017pbi})
\begin{equation}
K(z,\,z^{*})=|z|^2\,,\qquad\qquad W(z)=m_0(z+\beta)^2\,.
\label{1}
\end{equation}
The hidden sector of this model contains only one chiral superfield $z$.
When the parameter $\beta=\beta_0=-\sqrt{3}+2\sqrt{2}$,
the corresponding SUGRA scalar potential has two degenerate minima with zero energy density at the classical level.
One of them is a supersymmetric Minkowski minimum associated with $z^{(2)}=-\beta$.
In the other minimum of the SUGRA scalar
potential, $z^{(1)}=\sqrt{3}-\sqrt{2}$, local SUSY is broken and the gravitino becomes massive.
Thus this minimum can be identified
with the physical vacuum. In general, in SUGRA models an extra fine-tuning is needed
in order to arrange for the presence of
a supersymmetric Minkowski vacuum as well as for the degeneracy of the physical and second vacua.
This fine-tuning can be alleviated,
to some extent, within the no-scale inspired SUGRA models \cite{Froggatt:2017pbi}--\cite{Froggatt:2005nb}.

Since the vacuum energy density of supersymmetric states in flat Minkowski space is just zero, the cosmological constant
problem is thereby solved to first approximation by the assumed degeneracy of the physical and supersymmetric Minkowski vacua.
However, in the second vacuum SUSY can be broken dynamically in the observable sector. This results in an exponentially
suppressed energy contribution in the second phase, which is the only contribution to the vacuum energy density in this phase.
By virtue of the MPP, this energy density should be then transferred to the physical vacuum.
The analysis performed in the leading
one--loop approximation showed that the observed value of the cosmological constant can be reproduced in this case, even if the
SM gauge couplings are almost identical in both vacua \cite{Froggatt:2005nb}--\cite{Froggatt:2011fc}.

The aim of this note is to explore the dependence of the dark energy density on the parameters within the SUGRA models with
degenerate vacua more accurately. In particular, we use two--loop renormalization group (RG) equations to evaluate the scale
where the dynamical breakdown of SUSY takes place in the second phase. We argue that the observed value of the cosmological constant
can be reproduced when the SUSY breaking scale $M_S$ in the physical vacuum varies around $100\,\mbox{TeV}$.
This article is organized as follows. In the next section we briefly review the results of the numerical analysis
obtained before in the leading one--loop approximation. In section 3 we estimate the scale of the dynamical SUSY breaking
in the supersymmetric Minkowski vacuum using two--loop RG equations for different values of $M_S$.
The Higgsino dark matter scenario and the upper bound on $M_S$ in the simplest SUGRA models are discussed in section 4.
Our results are summarized in section 5.

\section{Estimation of the dark energy density in the one--loop approximation}

Hereafter we just assume that a phenomenologically viable SUGRA model with degenerate vacua
of the type discussed in the Introduction is realised in Nature. In other words there exist
at least two phases with exactly the same vacuum energy density. In the first phase, associated
with the physical vacuum, SUSY is broken and all sparticles have sufficiently large masses.
In the second phase the low energy limit of the theory under consideration is described
by a SUSY model in flat Minkowski space so that the corresponding vacuum energy density vanishes
in the leading approximation. Now we try to estimate the value of the energy density
in the second vacuum taking into account non--perturbative effects which may give rise to
the breakdown of SUSY in this phase at low energies\footnote{For detailed reviews regarding
the dynamical breakdown of SUSY see Ref.~\cite{22}}.

If the dynamical breakdown of SUSY takes place in the second phase, it should be caused by
the strong interactions in the observable sector. Indeed, the RG flow of the SM gauge couplings
$g_1$, $g_2$ and $g_3$, which are associated with $U(1)_Y$, $SU(2)_W$ and $SU(3)_C$ gauge
interactions respectively, obeys the RG equations (RGEs) that can be written to first order as
\begin{equation}
\frac{d\alpha_i(Q)}{dt}=\frac{b_i \alpha^2_i(Q)}{4\pi}\,,
\label{2}
\end{equation}
where $t=\log{Q^2}$, $Q$ is a renormalization scale, $i=1,2,3$ and $\alpha_i(Q)=g_i^2(Q)/(4\pi)$.
In the pure minimal supersymmetric standard model (MSSM) only the beta function of the strong gauge
coupling constant $\alpha_3(Q)$ exhibits asymptotically free behaviour, i.e. $b_3<0$. Therefore
$\alpha_3(Q)$ increases in the infrared region. When $\alpha_3(Q)$ becomes of the order of unity or
larger one can expect that the role of non--perturbative effects gets enhanced.

To simplify our analysis we restrict our consideration to the scenario where the values of the gauge
couplings at high energies are almost identical in the physical and supersymmetric Minkowski vacua.
Consequently the RG flow of these couplings down to the scale $M_S$ is also almost the same in both vacua.
Below the scale $M_S$ all superparticles in the physical vacuum decouple. As a result all beta functions change.
For instance, in the one--loop approximation the $SU(3)_C$ beta function $b_3=-3$ for $Q>M_S$ while below
$M_S$ it coincides with the corresponding SM beta function, i.e. $\tilde{b}_3=-7$. Because of this,
for $Q<M_S$ the RG running of $\alpha_3(Q)$ in the physical and supersymmetric Minkowski vacua ($\alpha^{(1)}_3(Q)$
and $\alpha^{(2)}_3(Q)$) is rather different. Using the matching condition $\alpha^{(2)}_3(M_S)=\alpha^{(1)}_3(M_S)$,
in the one--loop approximation one finds the value of the strong gauge coupling in the second phase \cite{Froggatt:2005nb}
\begin{equation}
\dfrac{1}{\alpha^{(2)}_3(M_S)}=\dfrac{1}{\alpha^{(1)}_3(M_Z)}-
\dfrac{\tilde{b}_3}{4\pi}\ln\dfrac{M^2_{S}}{M_Z^2}\, .
\label{3}
\end{equation}

In the supersymmetric Minkowski vacuum, all particles of the MSSM are massless. Therefore the RG flow of all
couplings is determined by the corresponding MSSM beta functions. At the scale $\Lambda_{c}$, where the supersymmetric
$SU(3)_C$ interactions become extremely strong in the second phase, the top quark Yukawa coupling $h^{(2)}_t(Q)$ is of
the same order of magnitude as the strong gauge coupling $g^{(2)}_3(Q)$. The large top quark Yukawa coupling is expected
to give rise to the formation of a quark condensate that breaks SUSY, resulting in a positive value of the vacuum
energy density \cite{Froggatt:2005nb}--\cite{Froggatt:2011fc}
\begin{equation}
\rho_{\Lambda} \sim \Lambda_{c}^4\, ,
\label{4}
\end{equation}
where in the one--loop approximation one obtains
\begin{equation}
\Lambda_{c}=M_{S}\exp\left[{\frac{2\pi}{b_3\alpha_3^{(2)}(M_{S})}}\right]\,.
\label{5}
\end{equation}
The induced vacuum energy density (cf. Eq.~\ref{5}) should be then interpreted
as the value of the cosmological constant
in the physical phase, by virtue of the postulated degeneracy of vacua.

From Eqs.~(\ref{3})--(\ref{5}) it follows that the cosmological constant and $\Lambda_{c}$ depend on the values
of $\alpha^{(1)}_3(M_Z)$ and the SUSY breaking scale, $M_S$, in the physical phase. Since in the one--loop
approximation $\tilde{b}_3 < b_3$, the QCD gauge coupling below $M_S$ is larger in the physical phase than
in the supersymmetric Minkowski vacuum. As a consequence, the value of $\Lambda_{c}$ is substantially lower than
the QCD scale $\Lambda_{QCD}$ in the SM and diminishes with increasing $M_S$. When $M_S$ is of the order of $1\,\mbox{TeV}$
and $\alpha^{(1)}_3(M_Z)\simeq 0.118$, one obtains $\Lambda_{c}\simeq 100\,\mbox{eV}$.
From the rough estimate of the energy
density (\ref{4}), it can be easily seen that the measured value of the cosmological constant is reproduced when
$\Lambda_{c}\sim 10^{-3}\,\mbox{eV}$.
In the framework of the MSSM the appropriate values of $\Lambda_{c}$ can be obtained
in the one--loop approximation only if $M_S\simeq 10^3-10^4\,\mbox{TeV}$ \cite{Froggatt:2003jm}--\cite{Froggatt:2017pbi},\cite{Froggatt:2005nb}--\cite{Froggatt:2011fc}.

In this approximation the measured dark energy density can also be reproduced if the MSSM particle content is supplemented
by an additional pair of $5+\bar{5}$ supermultiplets, which are fundamental and antifundamental representations of $SU(5)$ \cite{Froggatt:2005nb}--\cite{Froggatt:2011fc}.
The extra bosons and fermions would not affect gauge coupling unification very much, since they form complete representations
of $SU(5)$ (see for example \cite{29}). In the physical vacuum additional exotic states can gain masses around $M_S$. The corresponding
mass terms in the superpotential may be generated after the spontaneous breakdown of local SUSY because of the presence of
the bilinear terms $\left[\eta (5\cdot \overline{5})+h.c.\right]$ in the K$\Ddot{a}$hler potential of the
observable sector~\cite{30}.
In the supersymmetric Minkowski phase exotic states remain massless. As a result extra $5+\bar{5}$ supermultiplets give a considerable
contribution to the $\beta$ functions in the second vacuum. Indeed, the one--loop beta function of the $SU(3)_C$ interactions in the
second phase changes from $b_3=-3$ to $b_3=-2$. This reduces $\alpha^{(2)}_3(Q)$ and $\Lambda_{c}$ so that
in the one--loop approximation the observed value of the cosmological constant can be reproduced
even for $M_S\simeq 1\,\mbox{TeV}$
\cite{Froggatt:2005nb}--\cite{Froggatt:2011fc}.

\section{The results of the two--loop analysis}

In this context it is worthwhile examining how the estimates of the cosmological
constant and $\Lambda_{c}$ change when
the two--loop contributions to the beta functions are taken into account.
To simplify our analysis we ignore all Yukawa couplings
except that for the top quark. Then for $Q<M_S$ in the physical vacuum the RG flow of the gauge, top quark Yukawa and Higgs quartic couplings
is determined by the following set of RG equations of the SM (see for example \cite{schrempp}):
\begin{equation}
\begin{array}{rcl}
\dfrac{d\alpha_1}{dt} & = &\dfrac{\alpha_1^2}{4\pi}\Biggl[\dfrac{41}{10} + \dfrac{1}{4\pi}\Biggl( \dfrac{199}{50} \alpha_1
+ \dfrac{27}{10} \alpha_2 + \dfrac{44}{5} \alpha_3 - \dfrac{17}{10} Y_t \Biggr) \Biggr]\,,\qquad\qquad\qquad\qquad\qquad\qquad\\[4mm]
\dfrac{d\alpha_2}{dt} & = &\dfrac{\alpha_2^2}{4\pi}\Biggl[-\dfrac{19}{6} + \dfrac{1}{4\pi}\Biggl( \dfrac{9}{10} \alpha_1
+ \dfrac{35}{6} \alpha_2 + 12 \alpha_3 - \dfrac{3}{2} Y_t \Biggr) \Biggr]\,,\\[4mm]
\dfrac{d\alpha_3}{dt} & = &\dfrac{\alpha_3^2}{4\pi}\Biggl[-7 + \dfrac{1}{4\pi}\Biggl(\dfrac{11}{10} \alpha_1
+ \dfrac{9}{2} \alpha_2 - 26 \alpha_3 - 2 Y_t \Biggr) \Biggr]\,,
\end{array}
\label{6}
\end{equation}
$$
\begin{array}{rcl}
\dfrac{dY_t}{dt} & = & \dfrac{Y_t}{4\pi}\Biggl[ \dfrac{9}{2} Y_t - \dfrac{17}{20} \alpha_1 - \dfrac{9}{4} \alpha_2 - 8 \alpha_3
+ \dfrac{1}{4\pi}\Biggl(-12 Y_t^2 - 12 Y_t Y_{\lambda} + 6 Y_{\lambda}^2 + 36 \alpha_3 Y_t \\[0mm]
& + & \dfrac{225}{16} \alpha_2 Y_t + \dfrac{393}{80} \alpha_1 Y_t - 108 \alpha_3^2 - \dfrac{23}{4} \alpha_2^2 + \dfrac{1187}{600} \alpha_1^2
+ 9 \alpha_2 \alpha_3 + \dfrac{19}{15} \alpha_1 \alpha_3 - \dfrac{9}{20} \alpha_1 \alpha_2 \Biggr)\Biggr]\,,\\[0mm]
\dfrac{dY_{\lambda}}{dt} & = & \dfrac{1}{8\pi}\Biggl[ 24 Y_{\lambda}^2 + 12 Y_t Y_{\lambda} - 6 Y_t^2 - 9 \alpha_2 Y_{\lambda}
- \dfrac{9}{5} \alpha_1 Y_{\lambda} + \dfrac{27}{200} \alpha_1^2 + \dfrac{9}{20} \alpha_1 \alpha_2 + \dfrac{9}{8} \alpha_2^2\\[0mm]
& + & \dfrac{1}{4\pi}\Biggl( - 312 Y_{\lambda}^3 - 144 Y_t Y_{\lambda}^2 - 3 Y_t^2 Y_{\lambda} + 30 Y_t^3 + \dfrac{108}{5} \alpha_1 Y_{\lambda}^2
+ 108 \alpha_2 Y_{\lambda}^2 + \dfrac{17}{2} \alpha_1 Y_t Y_{\lambda}\\[0mm]
& + & \dfrac{45}{2} \alpha_2 Y_t Y_{\lambda} + 80 \alpha_3 Y_t Y_{\lambda}
- \dfrac{8}{5} \alpha_1 Y_t^2 - 32 \alpha_3 Y_t^2 + \dfrac{1887}{200} \alpha_1^2 Y_{\lambda}
+ \dfrac{117}{20} \alpha_1 \alpha_2 Y_{\lambda} - \dfrac{73}{8} \alpha_2^2 Y_{\lambda}\\[0mm]
& - & \dfrac{171}{100} \alpha_1^2 Y_t + \dfrac{63}{10} \alpha_1 \alpha_2 Y_t - \dfrac{9}{4} \alpha_2^2 Y_t
- \dfrac{3411}{2000} \alpha_1^3 + \dfrac{305}{16} \alpha_2^3 - \dfrac{1677}{400} \alpha_1^2 \alpha_2
- \dfrac{289}{80} \alpha_1 \alpha_2^2 \Biggr)\Biggr]\,,
\end{array}
$$
where $Y_t(Q)=h_t^2(Q)/(4\pi)$, $Y_{\lambda}(Q)=\lambda(Q)/(4\pi)$, whereas $h_t(Q)$ and $\lambda(Q)$ are the
top quark Yukawa and Higgs quartic couplings, respectively. The evolution of $g_i(Q)$, $h_t(Q)$ and $\lambda(Q)$
is computed for a given set of these couplings at the scale $Q=M_t$. The values of $g_i(M_t)$, $h_t(M_t)$ and
$\lambda(M_t)$ were specified in Ref.~\cite{Buttazzo:2013uya}. The results of our numerical analysis indicate that
the inclusion of two--loop corrections to the beta functions leads to minor variations of $g_i(M_S)$ and $h_t(M_S)$.
These variations are considerably less than a few percent.

Assuming that $\tan\beta\gg 1$, one can use the obtained values of $\alpha_i(M_S)$ and $Y_t(M_S)$ as well as
a set of two--loop RGEs
\begin{equation}
\begin{array}{rcl}
\dfrac{d\alpha_1}{dt} & = &\dfrac{\alpha_1^2}{4\pi}\Biggl[\dfrac{33}{5}+n+\dfrac{\alpha_1}{4\pi}\Biggl(\dfrac{199}{25}+\dfrac{7}{15}n\Biggr)
+ \dfrac{\alpha_2}{4\pi}\Biggl(\dfrac{27}{5}+\dfrac{9}{5}n\Biggr) + \dfrac{\alpha_3}{4\pi}\Biggl(\dfrac{88}{5}+\dfrac{32}{15}n\Biggr)
- \dfrac{26}{5} \dfrac{Y_t}{4\pi} \Biggr]\,,\\[4mm]
\dfrac{d\alpha_2}{dt} & = &\dfrac{\alpha_2^2}{4\pi}\Biggl[1+n+\dfrac{\alpha_1}{4\pi}\Biggl(\dfrac{9}{5}+\dfrac{3}{5}n\Biggr)
+ \dfrac{\alpha_2}{4\pi}\Biggl(25+7n\Biggr) + 24 \dfrac{\alpha_3}{4\pi} - 6 \dfrac{Y_t}{4\pi} \Biggr]\,,\\[4mm]
\dfrac{d\alpha_3}{dt} & = &\dfrac{\alpha_3^2}{4\pi}\Biggl[-3+n+\dfrac{\alpha_1}{4\pi}\Biggl(\dfrac{11}{5}+\dfrac{4}{15}n\Biggr)
+ 9 \dfrac{\alpha_2}{4\pi} + \dfrac{\alpha_3}{4\pi}\Biggl(14+\dfrac{34}{3}n\Biggr) - 4 \dfrac{Y_t}{4\pi} \Biggr]\,,\\[4mm]
\dfrac{dY_t}{dt} & = & \dfrac{Y_t}{4\pi}\Biggl[ 6 Y_t - \dfrac{13}{15} \alpha_1 - 3 \alpha_2 -\dfrac{16}{3} \alpha_3
+ \dfrac{1}{4\pi}\Biggl(-22 Y_t^2 + 16 \alpha_3 Y_t + 6 \alpha_2 Y_t + \dfrac{6}{5} \alpha_1 Y_t
\end{array}
\label{7}
\end{equation}
$$
\begin{array}{rcl}
& + & \Biggl(-\dfrac{16}{9}+ \dfrac{16}{3} n\Biggr)\alpha_3^2
+\Biggl(\dfrac{15}{2}+ 3 n\Biggr)\alpha_2^2 + \Biggl(\dfrac{2743}{450}+ \dfrac{13}{15} n\Biggr)\alpha_1^2
+ 8 \alpha_2 \alpha_3 \\[0mm]
& + & \dfrac{136}{45} \alpha_1 \alpha_3 + \alpha_1 \alpha_2 \Biggr)\Biggr]\,.
\end{array}
$$
to calculate the values of the gauge and top quark Yukawa couplings at the scale $M_X\simeq 2\cdot 10^{16}\,\mbox{GeV}$
in the physical vacuum\footnote{In SUSY extensions of the SM it is expected that near the scale $M_X$ the extended gauge symmetry
of the Grand Unified Theories is broken down to the SM gauge group}.
The parameter $n$ appearing in Eqs.~(\ref{7}) is the number of additional pairs of $5+\bar{5}$ supermultiplets
of $SU(5)$, that can survive to low energies together with the MSSM particle content.
In the case of the pure MSSM, i.e. $n=0$, the full set of two--loop RG equations is specified in Ref.~\cite{two--loop-rge}.
Here it is assumed that the values of $\tan\beta$ are sufficiently small, i.e. $\tan\beta\ll 50-60$, so that the $b$--quark
and $\tau$--lepton Yukawa couplings can be neglected to leading order.

\begin{figure}[h!]
\includegraphics[width=0.5\textwidth]{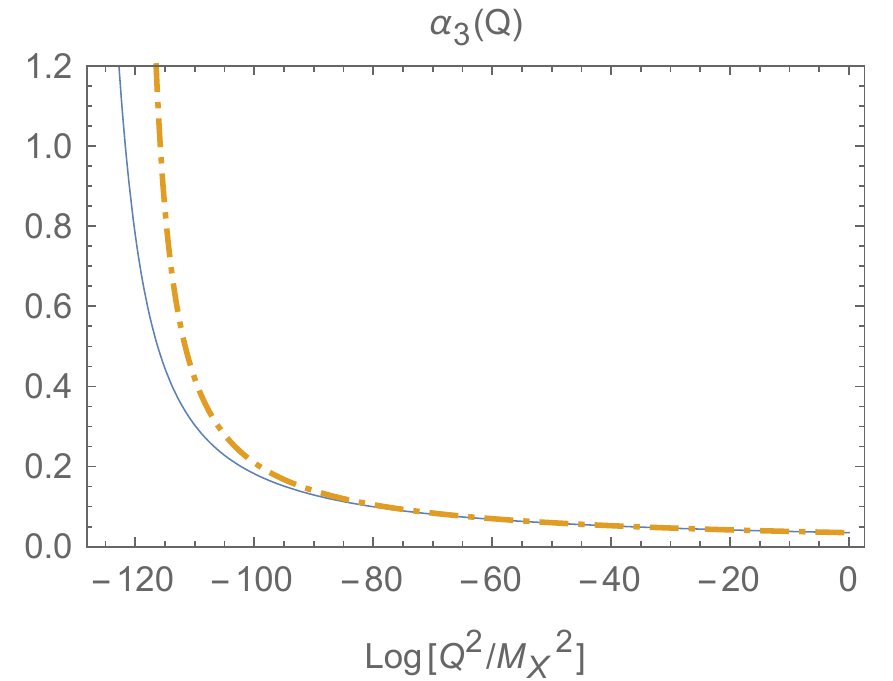}
\includegraphics[width=0.5\textwidth]{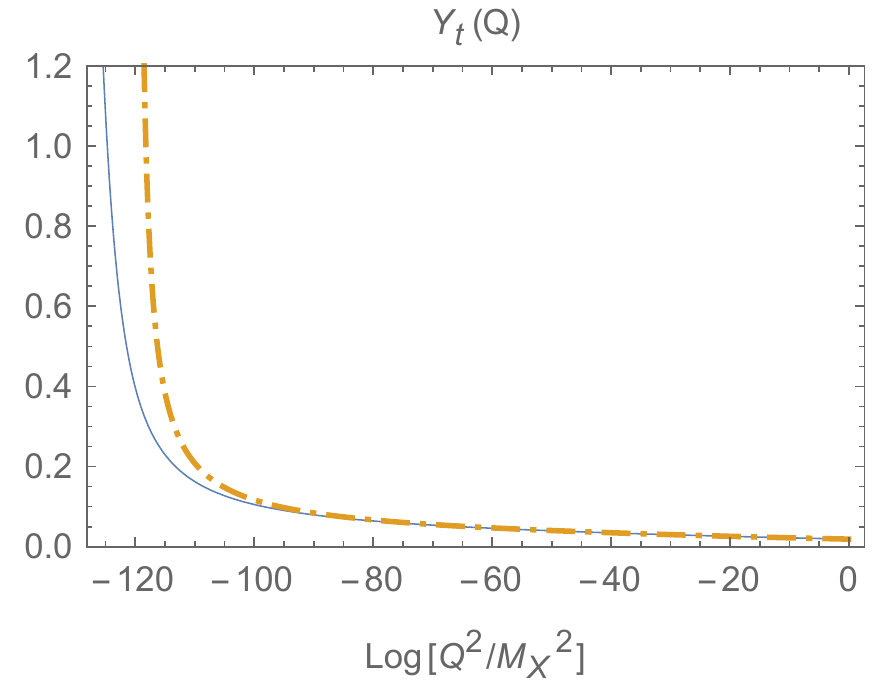}
\hspace*{3.5cm}{\bf (a)}\hspace*{8cm}{\bf (b) }\\
\caption{One--loop (dashed--dotted lines) and two--loop (solid lines) RG flow of couplings in the second vacuum
for $\alpha^{(2)}_i(M_X)=\alpha^{(1)}_i(M_X)$ and $Y^{(2)}_t(M_X)=Y^{(1)}_t(M_X)$:
{\it (a)} evolution of $\alpha_3(Q)$ from $M_X$ to low energies; {\it (b)} running of $Y_t(Q)$ from $M_X$ to low energies.
We set $M_S=100\,\mbox{TeV}$, $M_t=173.3\,\mbox{GeV}$ and $\alpha_3(M_Z)=0.118$ in the physical phase.}
\label{fig1}
\end{figure}
The computed values of $\alpha_i(M_X)$ and $Y_t(M_X)$ in the physical vacuum,
i.e. $\alpha^{(1)}_i(M_X)$ and $Y^{(1)}_t(M_X)$, allow one to estimate the position
of the Landau pole in the two--loop RG flow of $Y_t(Q)$ and $\alpha_3(Q)$ in the supersymmetric
phase, if the couplings at the scale $M_X$ in the first and second vacua are approximately the same
($\alpha^{(2)}_i(M_X)\approx \alpha^{(1)}_i(M_X)$ and $Y^{(2)}_t(M_X)\approx Y^{(1)}_t(M_X)$).
In the second vacuum $\alpha_i(Q)$ and $Y_t(Q)$ obey the two--loop RG equations (\ref{7}).
Although in our analysis we assume that the couplings at the scale $M_X$ in the physical and
second phases are almost identical, we still allow for the possibility that they might be slightly
different in these vacua. Since the one--loop analysis shows that $\Lambda_{c}$ does not change
when $\alpha^{(2)}_1(M_X)$ and $\alpha^{(2)}_2(M_X)$ vary we fix
\begin{equation}
\alpha^{(2)}_1(M_X)=\alpha^{(1)}_1(M_X)\,,\qquad \alpha^{(2)}_2(M_X)=\alpha^{(1)}_2(M_X)\,.
\label{8}
\end{equation}
At the same time the allowed range of variations of $\alpha^{(2)}_3(M_X)$ and $Y^{(2)}_t(M_X)$ is set by
\begin{equation}
\begin{array}{c}
0.97\cdot \alpha^{(1)}_3(M_X)\le \alpha^{(2)}_3(M_X) \le 1.03 \cdot \alpha^{(1)}_3(M_X)\,,\\[1mm]
0.97\cdot Y^{(1)}_t(M_X) \le Y^{(2)}_t(M_X) \le 1.03\cdot Y^{(1)}_t(M_X)\,.
\end{array}
\label{9}
\end{equation}
Because $\alpha^{(1)}_i(M_X)$ and $Y^{(1)}_t(M_X)$ are mainly determined by the SUSY breaking scale
in the physical vacuum, using Eqs.~(\ref{8})--(\ref{9}) one can find an interval of variations of
$\Lambda_{c}$ for each value of $M_S$.

The results of our numerical analysis are presented in Table~1 and Fig.~1.
In the second vacuum the two--loop corrections to the beta functions change the running of $Y_t(Q)$ and $\alpha_3(Q)$
considerably when these parameters are of the order of unity in the infrared region. To demonstrate this, let us consider
pure MSSM with $Y_t(Q)\simeq 0$. In this limit the Landau pole in the RG flow of $\alpha_3(Q)$ disappears. Moreover
when $Q\to 0$ the solutions of the two--loop RG equations (\ref{7}) are gathered near the infrared fixed point
\begin{equation}
\alpha_1(Q)\to 0\,,\qquad \alpha_2(Q)\to 0\,,\qquad \alpha_3(Q)\simeq \dfrac{6\pi}{7} \, .
\label{10}
\end{equation}
On the other hand, the sufficiently large values of the top quark Yukawa coupling associated with
the matching conditions (\ref{9}) result in the appearance of a Landau pole in the two--loop RG flow of $Y_t(Q)$ and
$\alpha_3(Q)$. Nevertheless, as one can see from Fig.~1, the inclusion of the two--loop corrections to the beta functions
reduces the growth of $\alpha_3(Q)$ and $Y_t(Q)$ in the infrared region. As a consequence the value of $\Lambda_{c}$
evaluated in the two--loop approximation is substantially lower than in the one-loop case (\ref{5}).
This is caused by the significant cancellation between the one--loop and two--loop contributions to the beta function
of the strong interaction for $\alpha_3(Q)\sim 1$. The value of $\Lambda_{c}$ grows with increasing
$\alpha^{(2)}_3(M_X)$, does not change much, when $Y^{(2)}_t(M_X)$ varies, and decreases with increasing $M_{S}$.
From the results presented in Table 1 it follows that in the two--loop approximation $\Lambda_{c}\simeq 0.001-0.002\,\mbox{eV}$
can be obtained for $M_S\simeq 20-400\,\mbox{TeV}$ when $\alpha^{(2)}_3(M_X)$ and $Y^{(2)}_t(M_X)$
change within the intervals given by Eq.~(\ref{9}).
\begin{table*}[ht]
\centering
\begin{tabular}{|c|c|c|c|c|}
\hline
$M_S\,(\mbox{TeV})$ & $10^4$ & $100$ & $20$ & $400$ \\
\hline\hline
$\Lambda_{c}\times 10^3\,(\mbox{eV})$ & $0.0002-0.012$ & $0.17-6.4$  & $1.8-63$     & $0.022-0.92$ \\
                                      & $(0.035-1.9)$  & $(27-1000)$ & $(290-9600)$ & $(3.7-150)$ \\
\hline
\end{tabular}
\caption{The intervals of variations of $\Lambda_{c}$ for different values of $M_S$.
The values of $\Lambda_{c}$ are computed using two--loop RG equations (\ref{7})
for $\alpha^{(2)}_i(M_X)$ and $Y^{(2)}_t(M_X)$, given by Eqs.~(\ref{8})-(\ref{9}) in which
$\alpha^{(1)}_i(M_X)$ and $Y^{(1)}_t(M_X)$ are calculated in the two--loop approximation
for each SUSY breaking scale, $M_S$, in the physical vacuum. We set $M_t=173.3\,\mbox{GeV}$
and $\alpha_3(M_Z)=0.118$ in the physical phase. The intervals of variations of $\Lambda_{c}$
calculated in the one--loop approximation are given in brackets.}
\label{L}
\end{table*}

If in the supersymmetric phase the low energy limit of the theory under consideration is described
by the MSSM with extra pairs of $5+\bar{5}$ supermultiplets of $SU(5)$, the Landau pole in the evolution
of $Y_t(Q)$ and $\alpha_3(Q)$ disappears entirely. For instance, when the MSSM particle spectrum is supplemented
by one additional pair of $5+\bar{5}$ supermultiplets, the solutions of the two--loop RG equations (\ref{7}) are
gathered near the infrared fixed point
\begin{equation}
\alpha_1(Q)\to 0\,,\qquad \alpha_2(Q)\to 0\,,\qquad \alpha_3(Q)\simeq 1.15\,,\qquad Y_t(Q)\simeq 1.01
\label{11}
\end{equation}
for $Q\to 0$. Thus a quark condensate may not get formed in this scenario.

\section{Higgsino dark matter and the upper bound on the SUSY breaking scale}

Recent experimental limits on the dark matter-nucleon scattering cross section obtained by
LUX \cite{Akerib:2016vxi}, PandaX-II \cite{Cui:2017nnn} and XENON1T \cite{Aprile:2018dbl}
indicate that within the MSSM the dark matter can be formed by the lightest neutralino,
which has to be either mostly higgsino or mostly wino. Since in the simplest SUSY extension of the SM,
namely the constrained version of the MSSM (cMSSM),
the lightest neutralino cannot be predominantly wino, here we focus
on the higgsino dark matter scenario\footnote{For a recent analysis see \cite{Athron:2016gor}.}. In the limit, when
the lightest SUSY particle (LSP) is almost pure higgsino, the cold dark matter relic density is basically
set by the mass of the LSP which is determined by the parameter $\mu$ in the MSSM superpotential, i.e. \cite{ArkaniHamed:2006mb}
\begin{equation}
\Omega_{\tilde{H}} h^2 \simeq 0.10 \left(\dfrac{\mu}{1\,\mbox{TeV}}\right)^2\,.
\label{12}
\end{equation}
Because the Planck observations lead to \cite{Ade:2015xua}
\begin{equation}
(\Omega h^2)_{\text{exp.}} = 0.1188 \pm 0.0010\,,
\label{13}
\end{equation}
the observed relic density of dark matter can be reproduced if $\mu$ is of the order of $1\,\mbox{TeV}$.

On the other hand, within the MSSM the parameter $\mu$ plays a key role in the EW symmetry breaking.
The sector responsible for the EW symmetry breaking in this extension of the SM involves two Higgs
doublets $H_1$ and $H_2$. The corresponding Higgs effective potential can be presented as a sum
\begin{eqnarray}
V=m_1^2|H_1|^2+m_2^2|H_2|^2-m_3^2(H_1 \epsilon H_2+h.c.)+
\sum_{a=1}^3 \dfrac{g_2^2}{8}\left(H_1^+\sigma_a H_1+H_2^+\sigma_a H_2\right)^2\nonumber\\
+\dfrac{{g'}^2}{8}\left(|H_1|^2-|H_2|^2\right)^2+\Delta V\,,\qquad\qquad\qquad\qquad\qquad
\label{14}
\end{eqnarray}
where $g'=\sqrt{3/5} g_1$. The last term, $\Delta V$, in Eq.~(\ref{14}) represents the contribution
of loop corrections to the Higgs effective potential. At the physical minimum of the scalar
potential (\ref{14}) the Higgs fields develop vacuum expectation values
\begin{equation}
<H_1>=\dfrac{1}{\sqrt{2}}\left(
\begin{array}{c}
v_1\\ 0
\end{array}
\right) , \qquad
<H_2>=\dfrac{1}{\sqrt{2}}\left(
\begin{array}{c}
0\\ v_2
\end{array}
\right)\,,
\label{15}
\end{equation}
breaking $SU(2)_W\times U(1)_Y$ gauge symmetry. Instead of $v_1$ and $v_2$, it is more convenient to use
$\tan\beta=v_2/v_1$ and $v=\sqrt{v_1^2+v_2^2} \approx 246\,\mbox{GeV}$. At the tree level ($\Delta V=0$)
the minimum conditions for the Higgs potential of  Eq.~(\ref{14}), i.e. $\dfrac{\partial V}{\partial v_1}=0$
and $\dfrac{\partial V}{\partial v_2}=0$, can be rewritten in the following form
\begin{eqnarray}
\dfrac{\bar{g}^2}{4}v^2=\dfrac{2(m_1^2-m_2^2\tan^2\beta)}{\tan^2\beta-1}\,,
\label{16}
\end{eqnarray}
\begin{eqnarray}
\sin 2\beta=\dfrac{2 m_3^2}{m_1^2+m_2^2}\,,
\label{17}
\end{eqnarray}
where $\bar{g}=\sqrt{g_2^2+g'^2}$.
When CP is conserved the Higgs spectrum of the MSSM contains two charged $H^{\pm}$, one pseudoscalar $A$
and two scalar ($h_1$ and $h_2$) states. If $M_S\gg M_Z$ the masses of the heaviest CP--even ($h_2$), charged and
CP--odd Higgs bosons are almost degenerate
\begin{eqnarray}
m^2_{h_2}\simeq m^2_{H^{\pm}}\simeq m_A^2\simeq \dfrac{2m_3^2}{\sin 2\beta}\sim M_S^2\,.
\label{18}
\end{eqnarray}
The mass of the lightest Higgs scalar in this limit is given by
\begin{eqnarray}
m_{h_1}\simeq \sqrt{M_Z^2 \cos^2 2\beta + \Delta}\,.
\label{19}
\end{eqnarray}
In Eq.~(\ref{19}) $\Delta$ denotes the loop corrections to $m^2_{h_1}$.

In order to get the lightest Higgs scalar with mass around $125\,\mbox{GeV}$ one should focus on
scenarios with $\tan\beta\gtrsim 10$. Nevertheless, even in this case a large loop contribution
of $\Delta \approx (85\,\mbox{GeV})^2$, which is nearly as large as $M_Z^2 \cos^2 2\beta$,
is needed to raise $m_{h_1}$ to $125\,\mbox{GeV}$. Such a loop contribution can be obtained if
$M_S\gtrsim 1\,\mbox{TeV}$. At the same time we restrict our consideration to $\tan\beta\lesssim m_t(M_t)/m_b(M_t)$,
i.e. $\tan\beta\lesssim 50-60$. Larger values of $\tan\beta$ result in the appearance of a Landau pole
that spoils the applicability of perturbation theory at high energies. In this range of $\tan\beta$
the flavour hierarchy problem may be partially solved because the $b$-quark and $\tau$-lepton Yukawa couplings
can be of the order of the top quark Yukawa coupling.

In SUGRA models $m_1^2=m_{H_1}^2+\mu^2$, $m_2^2=m_{H_2}^2+\mu^2$ and $m_3^2=|B||\mu|$, where
$m_{H_1}^2$, $m_{H_2}^2$ and $B$ are soft SUSY breaking terms. Within these models the SUSY breaking scale
$M_S$ is set by the gravitino mass $m_{3/2}$. It is also expected that $m_{H_1}^2\sim m_{H_2}^2\sim M_S^2$ and
$|\mu|\sim |B|\sim M_S$. However, in general this implies that the SUSY breaking scale should be of the order
of the EW scale and $\tan\beta\sim 1$. Such scenarios are basically ruled out by the LHC experiments.
Large values of $\tan\beta$ are induced when $\mu$ is substantially smaller than $M_S$ so that the LSP is predominantly
higgsino. In the phenomenologically acceptable scenarios of this type the parameter $m_{H_2}^2$ has to be adjusted
to ensure that the condition (\ref{16}) is fulfilled. Assuming that $m_{H_1}^2\sim M_S^2$ and using Eqs.~(\ref{17})--(\ref{18})
one finds the following estimate for the SUSY breaking scale
\begin{eqnarray}
M_S\sim c_B |\mu| \tan\beta\,,
\label{20}
\end{eqnarray}
where $c_B\simeq |B|/M_S$. Taking into account that $\tan\beta\lesssim 50-60$, $c_B$ is expected to be less than 10 and
$\mu\simeq 1\,\mbox{TeV}$, the SUSY breaking scale in this case should not exceed a few hundred $\mbox{TeV}$.
This is consistent with the prediction for $M_S$ obtained in section 3.

\section{Conclusions}

In this note we have examined the dependence of the cosmological constant on the SUSY breaking scale $M_S$
and other parameters within $N=1$ supergravity (SUGRA) models with degenerate vacua.
The corresponding SUGRA scalar potential has at least two exactly degenerate minima.
In the first (physical) vacuum local SUSY is broken in the hidden sector at the high energy scale,
inducing a non--zero gravitino mass $m_{3/2}$ and a set of soft SUSY breaking terms for the observable fields.
The tiny value of the cosmological constant in this physical phase
appears as a result of the fine-tuned, precise cancellation of different contributions.
In the second vacuum the low energy limit of the theory under consideration is described by a pure
SUSY model in flat Minkowski space. The breakdown of local SUSY in this supersymmetric Minkowski
phase is caused by the non--perturbative effects that give rise to the formation of the top quark
condensate near the scale $\Lambda_{c}$, where the $SU(3)_C$ gauge interactions in the observable sector
become strong, as well as inducing a non-zero and positive vacuum energy density $\sim \Lambda_{c}^4$.

In general an enormous fine tuning is required to get such phases with the same vacuum energy density.
Nevertheless, using a sufficiently accurate principle of these vacua being degenerate (MPP),
we adopt such a fine-tuning assumption for our estimation of the dark energy density.
In other words a small cosmological constant, which is computed in the second phase, is
transferred to the physical vacuum in which we live. Moreover we restrict our consideration
to the case where the gauge and top quark Yukawa couplings at the scale $M_X\simeq 2\cdot 10^{16}\,\mbox{GeV}$
are almost identical in both vacua. All other Yukawa couplings are neglected.
In our analysis we have used the two--loop RG equations of the SM to compute the evolution of couplings
in the physical phase below the SUSY breaking scale $M_S$ and the two--loop RG equations of the MSSM
to evaluate $\alpha^{(1)}_i(M_X)$ and $Y^{(1)}_t(M_X)$. Since we still allowed couplings to be slightly
different in the first and second phases, $\alpha^{(2)}_i(M_X)$ and $Y^{(2)}_t(M_X)$ are set by
Eqs.~(\ref{8})--(\ref{9}). Thus the values of $\alpha^{(2)}_1(M_X)$ and $\alpha^{(2)}_2(M_X)$
as well as the permitted ranges of variations of $\alpha^{(2)}_3(M_X)$ and $Y^{(2)}_t(M_X)$ are
fixed by the corresponding values of these couplings at the scale $M_X$ in the physical vacuum
which are mainly determined by the SUSY breaking scale $M_S$. Then we have explored the two--loop
RG flow of gauge and top quark Yukawa couplings in the second phase to estimate the position of
the Landau pole $\Lambda_{c}$. The numerical analysis carried out in the framework of the pure MSSM
has revealed that $\Lambda_{c}$ and the dark energy density in the second phase are mostly determined
by $\alpha^{(2)}_3(M_X)$ which decreases with increasing $M_{S}$. As a consequence $\Lambda_{c}$ becomes
much smaller than $0.001\,\mbox{eV}$ for $M_{S}\gtrsim 10^3\,\mbox{TeV}$. For the chosen interval
of variations of $\alpha^{(2)}_3(M_X)$ the values of $\Lambda_{c}\simeq 0.001-0.002\,\mbox{eV}$
can be obtained if $M_S$ changes from $20\,\mbox{TeV}$ to $400\,\mbox{TeV}$. We have also argued that
this prediction for the SUSY breaking scale is consistent with the upper bound on $M_S$ which can be
derived in the higgsino dark matter scenario within the simplest SUGRA models.

The prediction for $M_S$ found here suggests that most of the sparticles are too heavy
to be observed in the LHC experiments. When $M_S$ is sufficiently large, so that
$m_{3/2}\gtrsim 100\,\mbox{TeV}$, gravitinos decay before Big Bang Nucleosynthesis \cite{Ibe:2004tg}.
This allows one to avoid the gravitino problem \cite{gravitino-problem}.
If the effective SUSY breaking scale is close to $20\,\mbox{TeV}$ then some sparticles
(LSP, etc) can be much lighter than $M_S$ and therefore may be discovered at
either HE--LHC \cite{he-lhc} or FCC \cite{fcc}. In the near future the direct detection experiments
are going to constrain dark matter models including the higgsino dark matter scenario
in the MSSM with $M_S\gg 1\,\mbox{TeV}$ even further \cite{Aprile:2015uzo}.

\vspace{-5mm}
\section*{Acknowledgements}
\vspace{-3mm}
R.N. is grateful to S.~Bass, S.~I.~Blinnikov, E.~Boos, S.~Demidov, M.~Dvornikov, M.~Dubinin, S.~Duplij, S.~Godunov,
D.~Gorbunov, M.~Libanov, O.~Kancheli, N.~Krasnikov, D.~Kazakov, S.~F.~King, M.~M\"{u}hlleitner, V.~Rubakov, M.~Sher,
D.~Sutherland, S.~Troitsky, X.~Tata, M.~Vysotsky, A.~F.~Zakharov, E.~Zhemchugov and Yu.~Zinoviev for stimulating
discussions. H.B.N. thanks the Niels Bohr Institute for his emeritus status as well as Gia Dvali
for an invitation to Munich. C.F. thanks Glasgow University and the Niels Bohr Institute for hospitality and support.
C.F., H.N. and R.N. would like to thank L.~V.~Laperashvili for fruitful discussions.
R.N. and A.T. are grateful to G.~Taylor for very valuable comments regarding HE--LHC and FCC projects.
This research was funded by the University of Adelaide and the Australian Research Council through the
ARC Center of Excellence for Particle Physics at the Terascale (A.T.) CE110001104.









\end{document}